\documentstyle[11pt,aaspp4,epsfig]{article}   

\slugcomment{ApJ, accepted for publication 1999 July 8th}

\lefthead{}

\begin{document}

\def\spose#1{\hbox to 0pt{#1\hss}}
\def\lta{\mathrel{\spose{\lower 3pt\hbox{$\mathchar"218$}}
     \raise 2.0pt\hbox{$\mathchar"13C$}}}
\def\gta{\mathrel{\spose{\lower 3pt\hbox{$\mathchar"218$}}
     \raise 2.0pt\hbox{$\mathchar"13E$}}}
\def\Msun{{\rm M}_\odot}
\def\msun{{\rm M}_\odot}
\def\Rsun{{\rm R}_\odot}
\def\Lsun{{\rm L}_\odot}
\def\half{{1\over2}}
\def\RL{R_{\rm L}}
\def\zs{\zeta_{s}}
\def\zR{\zeta_{\rm R}}
\def\dJJ{{\dot J\over J}}
\def\dMM{{\dot M_2\over M_2}}
\def\tKH{t_{\rm KH}}
\def\eck#1{\left\lbrack #1 \right\rbrack}
\def\rund#1{\left( #1 \right)}
\def\wave#1{\left\lbrace #1 \right\rbrace}
\def\dd{{\rm d}}

\title{On the nature of XTE J0421+560/CI Cam}
\author{
        T.~Belloni\altaffilmark{1,2},
        S. Dieters\altaffilmark{3},
        M.E.~van den Ancker\altaffilmark{1},
        R.P. Fender\altaffilmark{1},
	D.W. Fox  \altaffilmark{4},
        B.A. Harmon \altaffilmark{5},
        M. van der Klis\altaffilmark{1},
        J.M. Kommers\altaffilmark{4},
	W.H.G. Lewin\altaffilmark{4},
        J. Van Paradijs\altaffilmark{1,6}
       }

\altaffiltext{1} {Astronomical Institute ``Anton Pannekoek'',
       University of Amsterdam and Center for High-Energy Astrophysics,
       Kruislaan 403, NL-1098 SJ Amsterdam, the Netherlands}
\altaffiltext{2}{Present address: Osservatorio Astronomico di Brera,
	Via E. Bianchi 46, I-23807 Merate (LC), Italy}
\altaffiltext{3} {Center for Space Plasma and Aeronomic Research, 
        University of Alabama in Huntsville, Huntsville, AL 35899, USA}
\altaffiltext{4} {Massachusetts Institute of Technology, Center for
	Space Research, Room 37-627, Cambridge, MA 02139, USA}
\altaffiltext{5} {NASA Marshall Space Flight Center, Huntsville, AL 35812, USA}
\altaffiltext{6} {Physics Department, University of Alabama in Huntsville,
	Huntsville, AL 35899, USA}

\begin{abstract}

We present the results of the analysis of RXTE, BATSE and optical/infrared
data of the 1998 outburst of the X-ray transient system XTE~J0421+560 (CI Cam).
The X-ray outburst shows a very fast decay (initial e-folding time
$\sim$0.5 days, slowing down to $\sim$2.3 days). The X-ray spectrum
in the 2-25 keV band is complex, softening considerably during decay
and with strongly variable intrinsic absorption.
A strong iron emission line is observed. No fast time variability
is detected ($<$0.5\% rms in the 1-4096 Hz band at the outburst peak).
The analysis of the optical/IR data suggests that the secondary is a B[e]
star surrounded by cool dust and places the system at a distance of 
$\gta$2 kpc. At this distance
the peak 2-25 keV luminosity is $\sim 4\times 10^{37}$ erg/s. 
We compare the properties of this peculiar system with those of the
Be/NS LMC transient A~0538-66 and suggest that CI Cam is of 
similar nature. The presence of strong radio emission during outburst
indicates that the compact object is likely to be a black hole or a
weakly magnetized neutron star.
\end{abstract}


\section{INTRODUCTION}

On 1998 March 31st, a bright X-ray transient was detected with the
All-Sky Monitor (ASM: \cite{lev96}) on board the Rossi X-ray 
Timing Explorer (RXTE)
(\cite{smi98}).  The source brightened quickly, reaching a maximum
brightness of $\sim$2 Crab a few hours after the discovery, then
decaying exponentially with a very short e-folding time (0.6 days,
\cite{rev98}). It was also detected with BeppoSAX (\cite{orl98}) and
(above 20 keV) with BATSE (\cite{pf98,hfp98}).  ASCA observations of
XTE~J0421+056 were made three days into the outburst (\cite{ue98}).
During the ASCA observations the source was fading with an e-folding time
of $\sim$ 30 hours and its spectrum was softening. The spectrum above
1 keV was fitted with a two-temperature (1.1 and 5.7 keV) optically
thin thermal model and an additional emission feature
at 6.4 keV. Interestingly, below 1 keV an additional component was
detected, which showed two flares during the observation.  Two
BeppoSAX observations of the transient confirmed the presence of a
very soft variable component in the X-ray spectrum (\cite{fro98,orr98}).
From the ASCA and the BeppoSAX data, it is evident that the X-ray spectrum 
between 1 and 10 keV is
essentially thermal, with the additional presence of a number of emission
lines (\cite{ue98,orr98}).

A transient radio counterpart of XTE J0421+560 was detected by
Hjellming \& Mioduszewski (1998\,a,b). The radio position coincides with
that of the variable star CI Cam ($=$ MWC 84) which had been
classified as a symbiotic star (\cite{ber95}).  This object, whose
visual magnitude is usually $V\sim 11.6$ (\cite{ber95}) had
brightened by $\gta$3.5 magnitudes in the R band (\cite{rob98}) and by
$>$2 magnitudes in B and V (\cite{gar98,hay98}).  The optical
outburst was also apparently very rapid. A comparison of the available
outburst photometry with 
historical photometry, shows that CI
Cam returned to within one magnitude from quiescence on April 6th, i.e.
the outburst lasted $\approx$ 6 days. 
The optical spectrum showed He II lines in addition to the complex emission
features observed in quiescence, but no indication of binarity was found
(\cite{ws98}).

In the radio, double S-shaped jets, moving at an apparent velocity of 0.15c
(for a distance of 1 kpc), have been reported (\cite{hm98b}).

Overall, almost all the observed properties of XTE~J0421+560/CI Cam
are rather peculiar and do not allow a straightforward classification of
the system beyond the assessment of the presence of a compact object,
given the large X-ray luminosity.

Here we present results of our analysis of the 
data collected during ten observations made with
the Proportional Counter Array (PCA) on board RXTE during eight
consecutive days following the decay from the peak. The BATSE data of the 
outburst are also presented.
In order to understand the nature of the system and estimate its distance,
we make use of near-infrared spectra and optical photometry obtained
shortly after the peak of the outburst.

\section{OBSERVATIONS}

\subsection{RXTE/PCA}

RXTE observed the source 10 times over roughly 8 days.  In order to
follow the spectral/timing evolution, we extracted PCA spectra for
each of the observation intervals in Table 1. For a description of the
instrument see Jahoda et al. (1996). The intervals were
chosen so that the flux decay during each interval was small in order
to minimize the effect of spectral/timing variations. During the
observation of April 3rd, the pointing direction was adjusted
following a better positional determination for the X-ray source.
The fluxes before the correction have been rescaled to take into account
the slight mispointing. 

The combined
PCA and ASM light curve is plotted in Fig. 1. After a rapid rise to
the peak in less than half a day (\cite{smi98}), the flux decayed
roughly exponentially with an e-folding time $\tau\sim$0.5 days for
two days. This value is consistent 
with the value of 0.56 d observed with BATSE (\cite{hfp98}),
although the energy range is different.
Afterwards, the decay slowed down, until after MJD 50908 it
was $\tau=2.34\pm0.05$ days. Notice that the intermediate decay time
scales reported with BeppoSAX (0.7-1.3 days, energy dependent, \cite{fro98}) 
and with ASCA (1.27 d, Ueda et al. 1998) come from the interval when the
flattening takes place (see Fig. 1), while the higher value obtained from the
second BeppoSAX pointing is consistent with our estimate.
The values reported by Revnivtsev, Emelyanov \& Borozdin (1998) from 
the same PCA data are consistent with our determination.

We searched the 1.2-3.5\,keV flux using the propane layer counts of the PCA
(April 1--3)
and found no deviations from a smooth decay comparable to those seen
by SAX and ASCA in the sub-keV light curve.
%

\subsection{BATSE hard X-ray observations}

XTE~J0421+056 was detected by the Burst and Transient Source Experiment
(BATSE) on board the Compton Gamma Ray Observatory (CGRO).
The BATSE Earth Occultation Technique was used to derive the flux and
spectra of the source in the energy range 20 keV--1 MeV. This technique allows
the flux of a bright hard X-ray
or gamma ray source to be sampled at times when the source rises above or
sets below the Earth's limb. 
For more details, see Fishman et al. (1989).

In the 20--100\,keV energy range covered by BATSE the light curve
shows a very rapid rise ($\sim$0.1\,day), peaking at March
31.98$\pm$0.075, followed by a rapid decline with an e-folding time which
matches that observed early-on with the PCA ($\sim$0.56\,day).

On the decline (MJD 50904 to 50905.0) the BATSE spectrum is well
modeled by a power law with photon number index of $-3.9\pm$0.3. In
comparison to other BHC transients this is a fairly soft spectrum.
The BATSE data are compatible with our fits to the 2--25\,keV data
(see Section 3.1).
There is marginal evidence for the hard X-ray spectrum being harder
during the rise and/or peak.

\subsection{Infrared}

Infrared spectroscopic observations of CI Cam were carried out across
the J-, H- and K-bands shortly after the peak of the outburst, on 1998
May 5 (J and K) and May 6 (H), at the UK Infrared Telescope
(UKIRT). The CGS4 instrument was used with a 0.6 arcsec slit width and
a spectral resolution of around 0.003 $\mu$m. A discussion of these
extremely rich emission line spectra, amongst the most detailed ever
obtained for an X-ray binary, can be found in Clark et al. (1998a),
in which 88 of more than 122 emission lines present in the spectrum
are identified.

\section{RESULTS}

\subsection{X-ray spectra}

The selection of data for our energy spectral analysis was done in
a standard way.  Only data well above the Earth's limb ($>$10$^\circ$),
with all 5 PCU's on, stable pointing, and full housekeeping were selected.  
 
The background spectra were estimated with the FTOOLS {\it pcabackest}
(V2.01) with the standard VLE$+$activation model for early (April 1st
to 5th) observations.  In the later observations the "faint'' (L7/240)
model was used for the background estimation.  The deadtime
correction factor was calculated for each background and source PHA
spectrum. The source PHA spectra were normalized by assuming that
there were no detectable source counts above 60\,keV. This was done to reduce
the effects of systematic errors in the estimated background.
Our fits were restricted to the 2--25\,kev range where the response of the PCA
is best determined.  To account for unknown systematic error a 1\% increase in
the formal errors  was applied to each PHA spectrum.

All fits included a component to take into account interstellar and intrinsic 
absorption from a cold absorber.  An emission line near 6.5\,keV is obvious 
in the raw PHA spectra of the early observations. Such a line was included in
all fits. Various simple models were used to model the continuum. Of these a
simple power law or blackbody type models did not fit the data, producing
unacceptably high $\chi^{2}$ values. A bremsstrahlung, broken power-law, a
cutoff power-law or a power-law with a high energy cutoff produced much better 
(though not formally acceptable) fits. The quality of these fits were similar.
We obtained results very similar to those of Revnivtsev et al. (1998) using the
absorbed cutoff power-law and line model. In all cases
except observations H and I better (90\% confidence using the F-test) fits
were obtained using a absorbed power-law with high-energy cutoff and line
model. The high-energy cutoff $HC$ is a multiplicative modification of a
power law of the form $HC=\exp((E_{cut}-E))/E_{fold})$ for $E>E_{cut}$ and
$HC=1$ for $E<E_{cut}$.

The fit results of this model are presented in Table 2. The timing history
of parameters showing significant variability is plotted in Fig. 2.
No matter which continuum model was used, the overall trends remain  unchanged.
It is evident that the spectrum softens with time; the power law steepens and
the folding energy moves from 11\,keV at peak down to near 6\,keV for the last
observation. Interestingly, the absorption increases from an initial (at peak
luminosity)  value near 2.2$\times 10^{22}$cm$^{-2}$ to near
3.8$\times 10^{22}$cm$^{-2}$, then decreases rapidly to become unmeasurable 
with the PCA. The actual values depend upon the model used. The initial
increase in N$_H$ is more dramatic when a cutoff power-law model is
used for the continuum model. Our N$_H$
measurements are not consistent with those from the contemporaneous SAX TOO1
observation, ours being considerably lower (see Orr et al. 1998). Notice that
the ASCA observation is also contemporaneous with the SAX TOO1 and our
I/J/K observations, and the value reported for N$_H$ is marginally consistent 
with our upper limits (Ueda et al. 1998). The values for SAX TOO2 are 
consistent with our upper limits (Orr et al. 1998). All we can conclude is
that the value of N$_H$ is strongly model-dependent, although it would be
difficult to interpret in this fashion the strong changes in absorption that 
we see between our early observations (A--G) and the later ones (H--R).
The line flux decreases as the overall luminosity and power-law flux decreases.
Initially the line flux is 4.8\% of the power-law flux, with a slight increase
to 5.5\% in observations E,F and G followed by a slow decrease to 2.7\% in the
last observation.

The line energies are all near 6.5\,keV in contrast to the near 6.7\,keV
measured using ASCA and SAX. This discrepancy is too large to be due to
uncertainties in the energy calibration of the PCA, but could be due to an
interaction between the Xenon edges of the response matrix and the
continuum/line model (Jahoda, priv. comm.(. We did find that the
energy of the iron line was  model dependent. In fact a simple power-law model
yielded a line energy near 6.7\,keV. The two SAX observations show no 
difference in
the line energy (Orr et al. 1998), in contrast to the shift claimed by
Revnivtsev et al. (1998) and evident in our fits with a cutoff power law
model and to a lesser extent with the power-law with high energy cutoff
model.  Since the continuum shape is changing markedly, we consider it likely
that the line shift is caused by the same coupling between response matrix and 
the model used to explain the lower energy of the iron line. With the more
complicated model of a cutoff power-law and edge we find no evidence for a
change in line energy.

With the above simple models, the residuals generally show a peak/dip structure 
in the 8--10\,keV range (Fig. 3). 
In some observations (i.e. H,I with the high energy
cutoff model) the residuals in the 15--25\,keV range dominate the final
$\chi^2$ values.  This indicates the need for an extra component in the
model of the PHA spectra. We found that the blackbody component of the
"standard'' BHC model  (disk-blackbody and cutoff power-law) did not improve
the fits and so was unnecessary. Adding an either an extra line near 8\,keV,
an edge or smeared edge  near 9.12\,keV (the energy expected for a iron
fluorescence line near 6.7\,keV; Nagase 1989),  or an extra power law
(specifically using a broken power-law instead of a power-law with the high
energy cutoff model) produced significantly better (95\% confidence) fits for
the earlier (before H) observations. 

The line energy of the extra line is near 8.3\,keV and we find no evidence for
a change in line energy. The line flux  also decreases approximately in step
with the power law component flux; fractionally about 0.39\% (lower in obs
H,I,J at 0.28\%). The edge or smeared edge component always has a small 
depth (MaxTau$\approx$0.1) and a fairly narrow width ($\approx$1\,keV).  For
the later observations an extra component was not warranted both by either the 
improvement in $\chi^2$ or the resulting best fit parameters i.e. the 8\,keV
line or edge became negligible, or the two power law slopes became the same.
An example of the fit residuals for various models is shown in Fig. 3 for
observation B (high flux, high absorption).

These four models are all equally good representations of the data. Thus we  
conclude that the spectra of CI Cam is more complex than the canonical 
X-ray transient, with no substantive evidence in the PCA data for moving lines
indicative of relativistic jets.

\subsection{X-ray variability}

To study the X-ray variability characteristics of XTE\,J0421+560 we made 
power density spectra for each of the individual data stretches listed in 
Table 1. 
Timing analyses show that the broad-band (1--4096 Hz) X-ray power
density spectra were flat throughout the outburst and probably
consistent (given the model uncertainties) with detector-modified
Poisson noise (\cite{zha96,mrg97}).
Thus the only excess X-ray variability exhibited by CI
Cam during its outburst decay was a smooth fading of
the source flux (which provides significant power at frequencies $\ll$1
Hz).

Upper limits (95\%-confidence) on the total 1--4096 Hz excess
variability are 0.5\% rms (dominated by systematic uncertainties) 
during the early (high count-rate)
observations, increasing to $\sim$10\% rms towards the end of the
outburst, with no detection registered at any time.
No coherent or quasiperiodic features were seen in the power spectra.
Upper limits on coherent pulsations (95\% confidence, 0.01--4096 Hz,
2--60 keV) are 0.1\% rms for the early observations, increasing to 2\%
by the end. In the restricted 5--12 keV band, limits increase from
0.1\% to 5\%.  Upper limits on QPO features are width-dependent;
increasing from 0.2\%, 0.4\%, and 0.7\% rms to 6\%, 10\%, and 15\% rms
over the course of the outburst decay for features 1/8, 1 and
8 Hz wide, respectively.

\section{The infrared/optical spectrum of CI Cam}

Comparing the quiescent optical/near-IR magnitudes listed in
Bergner et al. (1995) to the outburst values indicates that CI Cam had clearly 
brightened during the
outburst. In order to determine the nature of the bright mass donor
and the spectrum of the outburst component in the infrared, we first
fit the quiescent optical -- near-IR data. These data were fit by
means of a combined Kurucz plus optically thin dust (Waters, Cot\'e \&
Geballe 1988) model.  This fit to the optical/IR spectral energy distribution
is illustrated in Fig. 4; uncertainties in E(B--V) and
R$_{\rm V}$ (the ratio of total to selective extinction) are indicated
on the figure.  The data are consistent with a hot ($T_{\rm eff} \geq
22 000$K; i.e. B2 or earlier) star plus a large mass ($\geq 0.002
M_{\odot}$) of heated dust. Also indicated in Fig. 4 are the IRAS 12,
25 \& 60 $\mu$m fluxes, as well as the 100 $\mu$m upper limit; these
fit the dust model very well.  The value of $R_{\rm V}$ (3.7, compared
to the normal interstellar value of 3.1) indicates on average larger
particles in the dust shell, consistent with strong radiation pressure
having removed the smaller ($\leq 0.3 \mu$m; \cite{ste91}) particles.

The quiescent infrared data as well 
as the fact that both components in the optical/IR spectrum brighten during 
outburst clearly indicate that CI Cam is not a symbiotic system containing
a red giant, but that it contains a hot massive star
surrounded by a substantial dust shell. We conclude that these data
are more consistent with a B[e] classification (Hubert \& Jaschek 1997
and references therein) for the system. An LBV (Luminous Blue Variable,
see Humpreys \& Davidson 1994) interpretation is plausible,
but if this is the case it is has to be much more distant than 2 kpc, in 
contradiction with the upper limit on the distance obtained by
Clark et al. (1998b). 
We note that Lamers et al. (1997) list the system as being `unclassified /
compact planetary nebula ?'.
However, the IRAS 25/12 $\mu$m flux ratio of CI~Cam (0.35) places it
firmly outside the IRAS color-color box occupied by planetary nebulae
($F_{25}/F_{12} > 4.0$, see e.g. \cite{mir97}), excluding this
possibility.
Clark et al. (1998a) discuss the highly stratified and/or asymmetric
circumstellar environment of CI Cam, indicated by emission from both
high excitation HeII and low excitation NaI and CO-band emission;
this again is consistent with a B[e] model for the system.

In Fig. 4 we indicate VRI outburst photometry (from Clark et
al. 1998b) and our IR spectra, dereddened according to the anomalous
extinction law fitted to the quiescent data. After subtraction of the
quiescent spectra (dominated by the dust emission for $\lambda \geq 1
\mu$m) we find that the outburst excess across the JHK near--IR bands
is still very red. 
The cause of this red outburst component is not
clear; it may indicate heating of the dust during the outburst,
enhanced dust formation since the quiescent data of Bergner et
al. (1995), or a more exotic mechanism such as
high-frequency synchrotron emission. 
The optical -- infrared colors
are probably not compatible with additional quantities of new dust
(which would include small grains and hence a different value of
R$_{\rm V}$ from quiescence), and a synchrotron origin would have
implied flux densities in the radio regime which were far higher than
those observed, so we favor an explanation of significant, rapid
heating of the existing dust shell during the outburst to explain the
red color of the outburst component in the near--infrared.

Our model fits constrain the distance to CI Cam to be $\geq 350$
pc. The total A$_{\rm V}$ from our fit is $4.4 \pm 0.2$ mag. In the
direction of CI Cam, the local Orion spiral arm lies at a distance
of 200 -- 400 pc; the more distant Perseus spiral arm at around
2-4 kpc (Gilmore, King \& van der Kruit 1989). 
It seems quite plausible that CI Cam lies in the Perseus
spiral arm. At a distance of 2~kpc, 
its (quiescent) luminosity is around $10^5 L_{\odot}$,
while during outburst (assuming the same spectral distribution) it is
4 times higher.

Zorec (1997) has independently estimated the temperature, luminosity,
distance and reddening to CI Cam from the optical observations of
Bergner (1995). All the values derived (from optical quiescent
observations) are consistent with those arrived at in our independent
analysis, and we adopt his best distance estimate of 1750 pc to the
system. At this distance the mass donor is on or near the main
sequence with a mass of around 20M$_{\odot}$.  
As evidenced by the presence of high-excitation lines in the optical 
and near-infrared spectrum, during outburst the enhanced emission 
in the optical was dominated by hot gas. The brightening in the 
infrared during outburst is due to reprocessing of the enhanced
optical emission by the circumstellar dust.

\section{Discussion}

At the distance of $\sim 2$ kpc the peak X-ray luminosity of XTE\,J0421+560 
(several $10^{37}$ erg\,s$^{-1}$) is high enough that we can exclude 
accretion onto a white dwarf as the cause of the X-ray emission. Since the 
X-ray spectrum of XTE\,J0421+560 is not extremely soft, we can also exclude 
thermonuclear burning on a white-dwarf surface as the source of the X-ray 
emission (this would produce a supersoft X-ray source, see 
\cite{kah97}). The soft component detected by BeppoSAX and ASCA has
a spectrum similar to that of supersoft X-ray sources (see \cite{ue98}), 
but its relatively high temperature (0.12 keV) and relatively fast variability 
have not been observed among those sources.
We can also exclude a nova explosion: the X-ray luminosity would not
be inconsistent, but the X-ray spectrum is harder than in novae
and the optical spectrum does not have the features of a nova in 
outburst.

During the outburst the optical luminosity of CI Cam increased by an order 
of magnitude, on a time scale of order a day. It is plausible that this 
increase is caused by reprocessing of X rays. Such reprocessing dominates 
the optical emission of low-mass X-ray binaries (\cite{par95}), 
but in CI Cam the ratio of 
optical to X-ray luminosity outburst is much higher than in these objects, by 
about a factor of $10^3$. This indicates that the geometry of the 
reprocessing material in CI Cam is different from that in LMXB (in the 
latter it is an accretion disk).

To the best of our knowledge, the only object that shows properties 
similar to CI Cam, is the LMC transient A\,0538--66 (see Corbet et al.
1997 and references therein). The source was discovered with Ariel V by 
White \& Carpenter (1978), who observed two outbursts, each lasting 
several hours only, during which a peak luminosity close to 
$10^{39}$ erg\,s$^{-1}$ was reached. Additional brief (less than 12 hours) 
X-ray outbursts were detected by Johnston et al. (1980). 
Low-amplitude ($<$10$^{37}$erg/s) X-ray outbursts have also been detected
by Corbet et al. (1997) with ASCA and Mavromatakis \& Haberl (1993) with ROSAT.
The X-ray 
outbursts, which recur at the orbital period of 16.7 days (Johnston et al. 
1980) and have durations between a few hours and $\sim$10 days,
 are accompanied by optical outbursts, in which the brightness 
increases from a quiescent value $V = 14.7$ by up to 2.5 mag (see, e.g., 
\cite{den83}). The quiescent optical flux is dominated by the 
photospheric emission of the B2\,IIIe companion star. The optical outbursts 
last for several days (\cite{den83}). These outbursts 
recur during relatively short intervals (a few months) of activity, 
alternated by long (many years) periods of quiescence. Ponman, Skinner \&
Bedford (1984) found a strong anti-correlation between X-ray intensity and the
column density $N_{\rm H}$ for A\,0538--66. 
Mavromatakis \& Haberl (1993) report two outburst of different duration
from the ROSAT all-sky survey, both with mean luminosity of a few $10^{37}$
erg/s. Combining the ROSAT PSPC data with older Einstein MPC data,
they quantify the soft excess reported by Ponman, Skinner \& Bedford (1984):
they fit the spectrum with a cut-off power law with photon index 1.2 and
cutoff energy $\sim$11 keV, plus a soft component which can be described
by a blackbody with a temperature of 0.25 keV.

Thus, with respect to the brevity of the X-ray outbursts,
the large ratio of optical to X-ray 
luminosity, the X-ray spectrum and the irregular occurrence of activity periods 
CI Cam is very similar to A\,0538--66. It differs with respect to the 
relation between  
$N_{\rm H}$ and X-ray flux, to the frequency of the outbursts,
and to the physical properties of the 
circumstellar envelope of the companion, as reflected in their spectral 
classifications (B[e] for CI Cam, Be for A\,0538-66). 

A\,0538--66 is a member of the class of Be/X-ray transients, in which 
recurrent outbursts of X rays occur near periastron passage of the compact 
star in its eccentric orbit around the Be companion star. The compact star 
then enters the dense inner region of the slow equatorial wind of the Be 
star, part of which is captured and accreted at a rate that depends on the 
local density in the wind and the relative velocity of the neutron star 
through this wind. The shedding of mass by the Be star in its equatorial 
wind occurs at a strongly variable rate, which leads to finite trains of 
X-ray outbursts, separated by the orbital period (see, e.g., \cite{bil97}). 
The compact object in A\,0538--66 is a neutron star, as shown by 
the 69 ms pulsations in its X-ray intensity (\cite{ski82}). 
Apparently the solid 
angle of the Be star envelope intercepting X~rays is large enough that a 
substantial fraction of the X-ray luminosity is reprocessed into low-
energy photons, producing the soft component and the optical outburst.

The fact that for CI Cam the optical luminosity during outburst exceeds the 
X-ray luminosity (by a factor $>$10)
can be understood as a result of shielding: we do not see 
the primary X-ray emission but only a fraction which has been scattered by 
the material surrounding the compact object. This idea also provides a natural
explanation for the lack of variability at frequencies above 1 Hz: 
scattering-induced travel time delays of the observed photons have smeared 
out variability at these frequencies. The effect of such smearing has also 
been observed in Cyg X-3 (\cite{ber94}). 

In the case of A\,0538--66 the optical luminosity during outbursts is a 
substantial fraction (but less than unity) of the X-ray luminosity. Also, 
pulsations are detected. This indicates that in this source we see the 
primary X-ray emission in a relatively unobstructed fashion. The different 
spatial arrangement of the material surrounding the compact object in CI 
Cam relative to the line of sight to the observer (which may simply be a matter
of different inclination angle) provides a possible explanation for the 
different relations between $N_{\rm H}$ and the X-ray flux seen in CI Cam 
and A\,0538--66 (see Ponman, Skinner \& Bedford 1984). 

In summary, although the nature of the compact star in CI Cam has not been 
revealed by the observations presented here, the properties of CI Cam as 
seen during its 1998 outburst suggest it is a B[e]/X-ray binary. 
In all Be/X-ray binaries for which we have definite information about the
nature of the compact star, this is a neutron star (i.e. pulsator).
However, CI Cam
has been detected as a relatively bright radio source (\cite{hm98}). 
The detection of radio emission at all would seem to exclude a high
magnetic field neutron star (X-ray pulsar) from the observed strong
anticorrelation between the two properties (Fender et al. 1997). The
strength of the radio emission at the peak of the outburst ($\sim$1 Jy at cm
wavelengths) is more typical of black hole candidates such as GRO~J1655-40
than of low magnetic field neutron star transients such as Aql X-1 (typically
around a few mJy), although several strong radio sources e.g. SS 433, Cyg
X-3, LSI+61 303 may contain neutron stars (see \cite{hh95,fbw97}).
In conclusion, our results suggest that XTE~J0421+560/CI~Cam is a peculiar
system consisting of a B[e] star and either a low magnetic field neutron
star or a black hole. In either case it would be the first system of
its kind.

\acknowledgements

MvdA acknowledges financial support from NWO/NFRA grant 781-76-015.
RPF thanks Simon Clark for useful discussions and Tom Geballe for obtaining and
reducing the IR spectra, and was supported by EC Marie Curie Fellowship
ERBFMBICT~972436. SD and JVP acknowledge the support of LTSA grant number 
NAG~5-6021.

{}

\newpage

\clearpage

\begin{deluxetable}{llccccc}
\footnotesize
\tablecaption{Observation log. Times are in UT, exposures in seconds.
	the MJD values refer to the start of the observation.}
\tablewidth{0pt}
\tablehead{
\colhead{ID}    &
\colhead{Date}  &
\colhead{Start}   &
\colhead{End}   &
\colhead{Exp.}  &
\colhead{MJD}
}
\startdata
A&1998 Apr  1 &  1:55 &  2:14 & 1028 & 50904.0798\nl
B&1998 Apr  1 &  2:14 &  2:32 & 1014 & 50904.0930\nl
C&1998 Apr  1 &  3:38 &  3:50 &  667 & 50904.1514\nl
D&1998 Apr  1 &  6:41 &  7:10 & 1597 & 50904.2784\nl
E&1998 Apr  1 &  8:20 &  9:14 & 2484 & 50904.3472\nl
F&1998 Apr  1 &  9:56 & 10:10 &  769 & 50904.4139\nl
G&1998 Apr  1 & 10:29 & 10:50 & 1199 & 50904.4368\nl
H&1998 Apr  2 &  6:40 & 10:47 & 8023 & 50905.2777\nl
I&1998 Apr  3 &  5:05 &  5:51 & 2063 & 50906.2118\nl
J&1998 Apr  3 &  5:53 &  9:09 & 5219 & 50906.2451\nl
K&1998 Apr  3 & 11:37 & 11:49 &  701 & 50906.4840\nl
L&1998 Apr  4 &  6:40 &  9:14 & 3144 & 50907.2777\nl
M&1998 Apr  5 &  6:39 & 10:08 & 7805 & 50908.2770\nl
N&1998 Apr  6 &  3:27 &  4:28 & 1502 & 50909.1437\nl
O&1998 Apr  7 &  2:36 &  4:26 & 4180 & 50910.1083\nl
P&1998 Apr  7 & 18:08 & 18:22 &  743 & 50910.7555\nl
Q&1998 Apr  8 &  3:45 &  6:02 & 5888 & 50911.1562\nl
R&1998 Apr  9 &  6:38 &  7:30 & 3084 & 50912.2764\nl
\enddata
\end{deluxetable}

\begin{deluxetable}{lccccccccccc}
\scriptsize
\tablecaption{Spectral parameters for the fits to the RXTE/PCA data 
	described in the text (power-law+gaussian line, modified
	by a high-energy cutoff). 
        Errors are $1\sigma$. Reduced $\chi^2$ are for 47 degrees
        of freedom.
        Observation P was too short to obtain meaningful values.}
\tablewidth{0pt}
\tablehead{
  \colhead{ID}		&
  \colhead{Rate}	&
  \colhead{N$_H$}	&
  \colhead{$\Gamma$}	&
  \colhead{E$_l$}	&
  \colhead{W$_l$}	&
  \colhead{EQW$_l$}	&
  \colhead{E$_{cut}$}	&
  \colhead{E$_{fold}$}	&
  \colhead{F$_{pl}$}	&
  \colhead{F$_l$}	&
  \colhead{$\chi^2$} \\
  \colhead{  }		&
  \colhead{cts/s}	&
  \colhead{cm$^{-2}$}	&
  \colhead{        }	&
  \colhead{keV}  	&
  \colhead{keV}  	&
  \colhead{eV}  	&
  \colhead{keV} 	&
  \colhead{keV} 	&
  \colhead{erg/cm$^{2}$/s}	&
  \colhead{erg/cm$^{2}$/s}	&
  \colhead{        }\\
}
\startdata
A&32314&2.20$\pm$0.36&1.42$\pm$ 0.09      &6.57$\pm$0.03&0.49$\pm$0.03&642&
  5.20$\pm$0.46       &11.23$\pm$ 0.75      & 9.63$\times 10^{-8 }$&
  4.66$\times 10^{-09}$&0.84\nl
B&31596&2.43$\pm$0.36&1.49$\pm$ 0.09      &6.54$\pm$0.03&0.47$\pm$0.06&608&
  4.47$\pm$0.46       &11.68$\pm$ 0.75      & 9.55$\times 10^{-8 }$&
  4.23$\times 10^{-09}$&0.88\nl
C&27217&2.76$\pm$0.38&1.54$\pm$ 0.10      &6.55$\pm$0.03&0.34$\pm$0.18&637&
  5.44$\pm$0.48       &11.51$\pm$ 0.79      & 8.26$\times 10^{-8 }$&
  3.98$\times 10^{-09}$&0.94\nl
D&23297&3.51$\pm$0.36&1.57$\pm$ 0.08      &6.53$\pm$0.03&0.30$\pm$0.04&645&
  5.56$\pm$0.45       &12.27$\pm$ 0.81      & 7.29$\times 10^{-8 }$&
  3.53$\times 10^{-09}$&0.98\nl
E&19898&3.76$\pm$0.36&1.53$\pm$ 0.08      &6.53$\pm$0.03&0.48$\pm$0.06&697&
  5.37$\pm$0.42       &11.38$\pm$ 0.72      & 6.21$\times 10^{-8 }$&
  3.29$\times 10^{-09}$&1.10\nl
F&17325&3.70$\pm$0.39&1.51$\pm$ 0.09      &6.52$\pm$0.03&0.25$\pm$0.03&719&
  5.47$\pm$0.42       &10.96$\pm$ 0.74      & 5.36$\times 10^{-8 }$&
  2.95$\times 10^{-09}$&1.33\nl
G&16386&3.69$\pm$0.37&1.48$\pm$ 0.09      &6.53$\pm$0.03&0.24$\pm$0.03&721&
  5.46$\pm$0.39       &10.90$\pm$ 0.65      & 5.07$\times 10^{-8 }$&
  2.81$\times 10^{-09}$&1.18\nl
H& 3225&$<0.082$     &1.54$\pm$ 0.04      &6.55$\pm$0.03&0.32$\pm$0.03&675&
  5.83$\pm$0.22       &10.09$\pm$ 0.22      & 8.63$\times 10^{-9 }$&
  4.62$\times 10^{-10}$&0.95\nl
I&  947&$<0.23$      &2.12$_{-0.03}^{+0.06}$&6.53$\pm$0.03&0.29$\pm$0.06&636&
  6.83$_{-0.31}^{+0.40}$&12.39$_{-0.49}^{+0.83}$& 2.58$\times 10^{-9 }$&
  1.22$\times 10^{-10}$&1.42\nl
J&  810&$<0.15$      &2.11$_{-0.03}^{+0.04}$&6.58$\pm$0.03&0.27$\pm$0.06&622&
  6.44$_{-0.27}^{+0.39}$&12.16$_{-0.39}^{+0.52}$& 2.20$\times 10^{-9 }$&
  1.01$\times 10^{-10}$&1.47\nl
K&  553&$<0.59$&2.25$_{-0.29}^{+0.15}$&6.55$\pm$0.04&0.29$\pm$0.08&651&
  7.02$_{-0.40}^{+1.02}$&13.15$_{-0.71}^{+2.71}$& 1.52$\times 10^{-9 }$&
  7.00$\times 10^{-11}$&0.86\nl
L&  288&$<0.57$&2.46$_{-0.04}^{+0.15}$&6.58$\pm$0.05&0.33$\pm$0.06&673&
  6.65$_{-0.86}^{+0.40}$&12.27$_{-0.66}^{+2.30}$& 8.39$\times 10^{-10}$&
  3.64$\times 10^{-11}$&0.97\nl
M&  138&$<0.24$      &2.69$_{-0.03}^{+0.07}$&6.56$\pm$0.03&0.27$\pm$0.07&649&
  7.18$_{-0.27}^{+0.34}$& 9.78$_{-0.59}^{+0.80}$& 3.92$\times 10^{-10}$&
  1.60$\times 10^{-11}$&1.19\nl
N&   86&$<0.16$      &2.79$\pm$ 0.05      &6.57$\pm$0.04&$<$0.26      &655&
  7.30$\pm$0.48       & 8.10$_{-1.15}^{+1.50}$& 2.48$\times 10^{-10}$&
  9.77$\times 10^{-12}$&1.43\nl
O&   57&$<0.25$      &2.92$_{-0.05}^{+0.08}$&6.55$\pm$0.04&$<$0.32      &661&
  7.50$\pm$0.41       & 6.84$_{-0.93}^{+1.19}$& 1.68$\times 10^{-10}$&
  6.43$\times 10^{-12}$&0.66\nl
Q&   41&$<0.24$      &2.98$_{-0.04}^{+0.07}$&6.63$\pm$0.07&$<$0.35      &522&
  6.96$_{-0.65}^{+0.45}$& 6.96$_{-1.22}^{+1.82}$& 1.25$\times 10^{-10}$&
  3.58$\times 10^{-12}$&0.98\nl
R&   30&$<1.09$      &3.10$_{-0.09}^{+0.26}$&6.56$\pm$0.09&$<$0.29      &490&
  7.33$_{-0.85}^{+0.63}$& 6.25$_{-1.55}^{+2.59}$& 9.39$\times 10^{-11}$&
  2.51$\times 10^{-12}$&0.51\nl
\enddata
\end{deluxetable}

\newpage

\begin{figure}
\centerline{\psfig{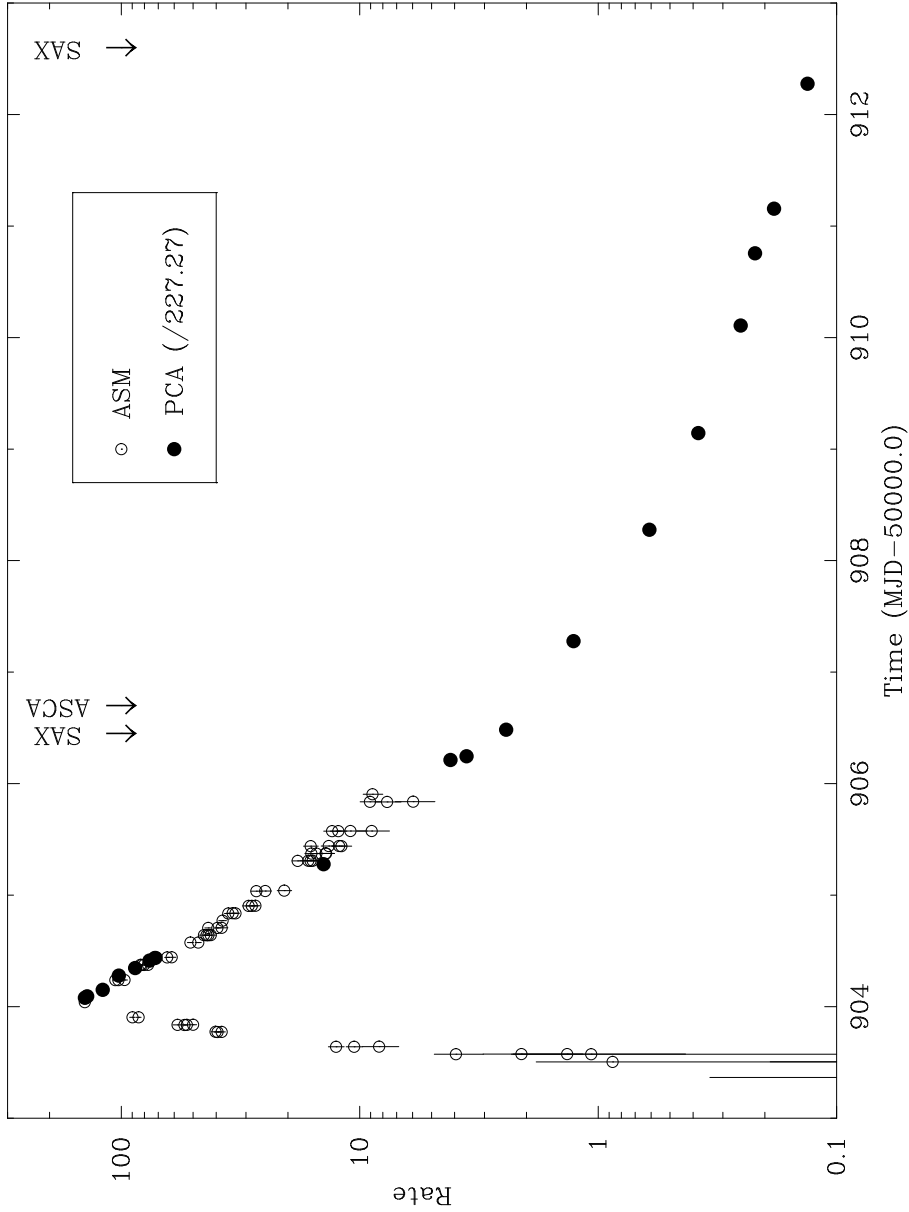}}
\caption{
Combined ASM/PCA light curve of XTE~J0421+560. The ASM points
        are from http://space.mit.edu/XTE/ASM\_lc.html, the PCA points are from
        this work. The PCA rate has been renormalized to match the ASM rate
        at the peak. The arrows indicate the middle of the ASCA and BeppoSAX
        observations.
}
\end{figure}

\begin{figure}
\centerline{\psfig{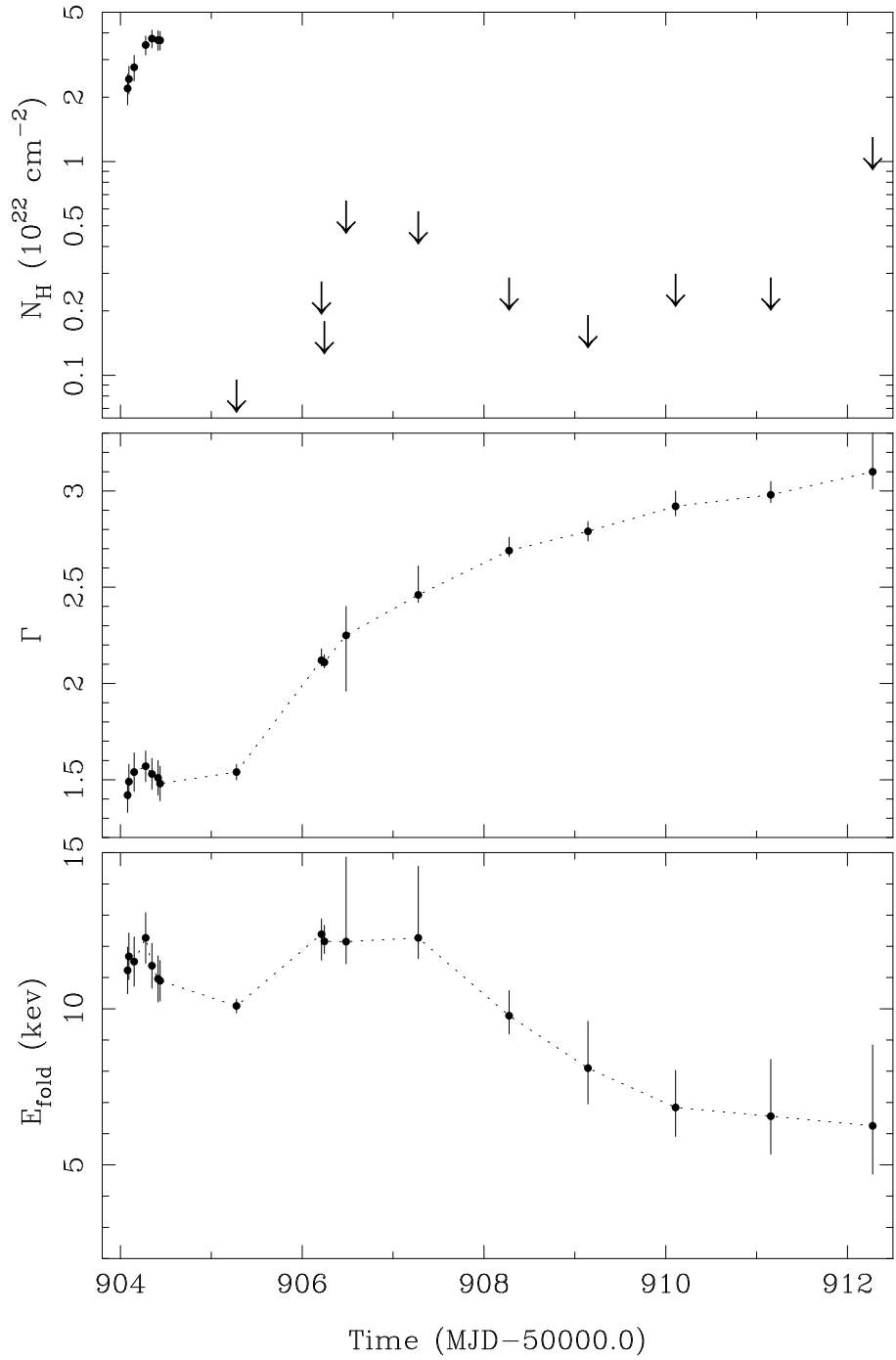}}
\caption{
Time history of variable parameters in the X-ray spectrum of
        XTE~J0421+560. Top panel: absorption; middle panel: power law
        photon index; bottom panel: cutoff energy.
}
\end{figure}

\begin{figure}
\centerline{\psfig{file=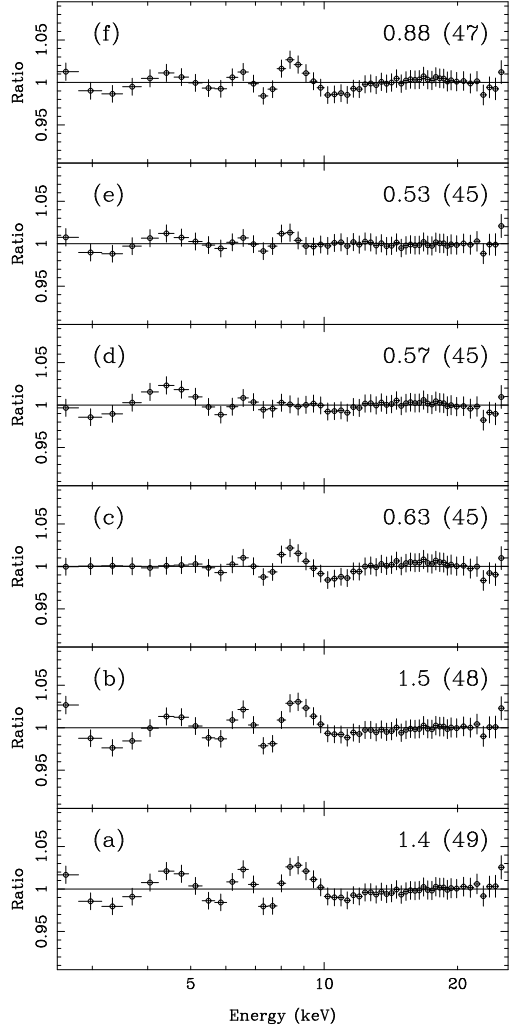, width=12cm}}
\caption{
Ratio data/model for the spectrum of observation B for different
        spectral models described in the text. The numbers in each panel
        are reduced $\chi^2$ and number of degrees of freedom. The models
        are:
        (a) bremsstrahlung plus gaussian line;
        (b) cutoff power law plus gaussian line;
        (c) broken power plus gaussian line;
        (d) power law with high energy cutoff plus two gaussian lines;
        (e) power law with high energy cutoff plus gaussian line
            and absorption edge;
        (f) power law with high energy cutoff plus gaussian line.
}
\end{figure}

\begin{figure}
\centerline{\psfig{file=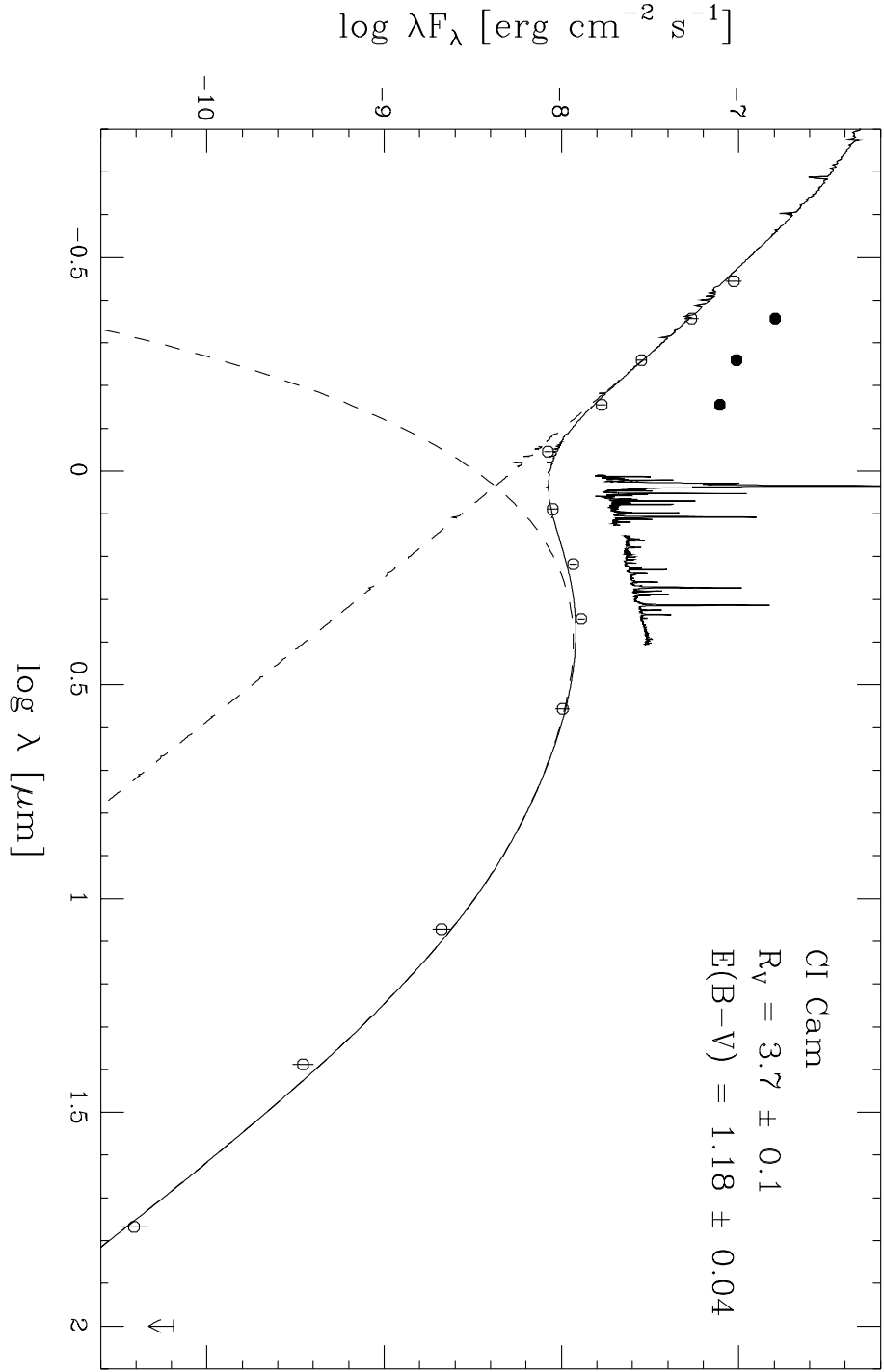, width=12cm}}
\caption{
Fit to the observed quiescent optical and near-IR photometry
(open symbols) of CI Cam with a Kurucz hot star + optically thin dust
(dashed lines; combined model solid line) and anomalous reddening
model. Also indicated are the IRAS 12, 25 and 60 $\mu$m flux densities
(plus the 100$\mu$m upper limit), which are consistent with the dust
model. Also indicated are optical photometry and infrared spectra
obtained during the outburst. Subtraction of the quiescent emission
reveals the outburst component in the near--IR to be very red,
indicative of rapid heating of the dust shell during the outburst.
}
\end{figure}

\end{document}